\documentclass[prl,floatfix,showpacs,preprintnumbers,amsmath,amssymb,superscriptaddress]{revtex4}
\usepackage{graphicx,psfrag}
\usepackage{amsfonts}

\begin{document}
 \title{Three-dimensional massive gravity  and the bigravity black hole }
 \author{M\'{a}ximo Ba\~{n}ados\footnote{J.S. Guggenheim Memorial Foundation fellow.}}
 \affiliation{  Physics Department, University of Oxford, Oxford, OX1 3RH, UK}
\affiliation{Departamento de F\'{\i}sica, P. Universidad Cat\'{o}lica de Chile, Casilla 306, Santiago, Chile.}
\author{ Stefan Theisen}
\affiliation{ Max-Planck-Institut f\"ur Gravitationsphysik,
Albert-Einstein-Institut, 14476 Golm, Germany}

\begin{abstract}
We study three-dimensional massive gravity formulated as a theory with two dynamical metrics, like the $f$-$g$ theories of Isham-Salam and Strathdee. The action is parity preserving and has no higher derivative terms. The spectrum contains a single massive graviton. This theory has several features discussed recently in TMG and NMG. We find warped black holes, a critical point, and generalized Brown-Henneaux boundary conditions.
\end{abstract}

\maketitle

\section{Massive 3d gravity}

Massive three dimensional gravity \cite{TMG} has recently been under great scrutiny. It was shown in \cite{LISongStrominger} that topologically massive gravity (TMG) with a cosmological constant (fine tuned with the Chern-Simons coupling) yields a theory with interesting properties. See \cite{CD,Grumiller,CD2,V} for subsequent work.  TMG is parity violation and propagates only one helicity of the massive graviton.  A different parity-preserving formulation for massive gravity, called New Massive Gravity (NMG), has been put forward in \cite{BHT,Bergshoeff:2009aq} involving a higher derivative action which nevertheless carries no ghost.

It is the purpose of this note to display yet another action for a single massive graviton in three dimensions. This formulation, known as {\it f-g theory}, has actually been known for a long time \cite{ISS}. The idea is to consider two spin-2 fields $g_{\mu\nu}$ and $f_{\mu\nu}$, coupled by a potential $U(g,f)$,
\begin{equation}\label{I}
I[g_{\mu\nu},f_{\mu\nu}] = {1 \over 16\pi G} \int \left[ \sqrt{-g} \left( R(g)+ {2 \over \ell^2} \right) + \sigma\sqrt{-f} \left( R(f)+ {2 \over \ell^2 } \right)  - U(g,f)  \right]
\end{equation}
where $\sigma$ is a dimensionless constant. The action (\ref{I}) is parity preserving and has no higher derivative terms. This action is invariant under diagonal diffeomorphisms acting on both metrics. Then, the spectrum contains a massless graviton (for any potential $U$). What is special in three dimensions is that the massless mode is trivial (up to boundary effects discussed below) and can be discarded. The propagating mode is then a single massive spin-2 field with 2 (linearized) degrees of freedom, just as NMG \cite{BHT}.

There is much freedom in the choice of potential. For interactions not involving derivatives a classification can be given \cite{Damour:2002ws}. As working example, we  consider here the original Isham-Salam-Strathdee potential \cite{ISS} giving rise to a Pauli-Fierz theory,
\begin{equation}\label{}
U(g,f)= {\nu \over \ell^2} \sqrt{-f} (g_{\mu\nu}-f_{\mu\nu})(g_{\alpha\beta}-f_{\alpha\beta})(f^{\mu\alpha}f^{\nu\beta} - f^{\mu\nu}f^{\alpha\beta}  ).
\end{equation}
Here $f^{\mu\nu}$ represents the inverse of $f_{\alpha\beta}$.  The action (\ref{I}) contains three parameters. $\sigma$ and $\nu$ are dimensionless while $\ell$ is a length.  The volume element in the potential $\sqrt{-f}$ can be generalized to $|-g|^u |-f|^{{1 \over 2}-u}$. As far as the linear theory is concerned (on the background with $f_{\mu\nu}=g_{\mu\nu}$), all choices of $u$ give the same theory. At the non-linear level the theories may be very different. We have chosen $u=0$ motivated by phenomenological applications of bigravity in 4 dimensions\cite{BFS,BGRS,B2}.

Massive gravity theories suffer from the Boulware Deser instability \cite{BD,BD2}, at non-linear level. In this work we shall mostly consider the linear theory.

\section{Linear Theory}

The action (\ref{I}) has a natural (this is not the only AdS solution, see below) AdS background
\begin{equation}\label{bu}
g_{\mu\nu}=f_{\mu\nu}=\bar g_{\mu\nu}, \ \ \ \  \mbox{with} \ \ \ \ \ \   \bar g_{\mu\nu}dx^\mu dx^\nu = -\left( 1 + {r^2 \over \ell^2} \right) dt^2 + {dr^2 \over 1 + {r^2 \over \ell^2}} + r^2 d\phi^2
\end{equation}
Note that the potential plays no role in this solution because $U$ and its derivatives vanish at $g_{\mu\nu}=f_{\mu\nu}$.

Now consider fluctuations $h_{\mu\nu}$ and $\rho_{\mu\nu}$ defined by,
\begin{equation}\label{}
g_{\mu\nu} = \bar g_{\mu\nu} + h_{\mu\nu}, \ \ \ \ \ f_{\mu\nu} = \bar g_{\mu\nu} + \rho_{\mu\nu}.
\end{equation}
The action for the fluctuations becomes
\begin{equation}\label{I1}
I[h_{\mu\nu},\rho_{\alpha\beta}] = {1 \over 16\pi G} \int \sqrt{-\bar g}\left( h^{\mu\nu} ({\cal G} h)_{\mu\nu} + \sigma\, \rho^{\mu\nu} ({\cal G} \rho)_{\mu\nu} - {\nu \over \ell^2} (h-\rho) \cdot (h-\rho) \right)
\end{equation}
where ${\cal G}$ is the Pauli-Fierz operator on curved AdS,
\begin{eqnarray}\label{}
h^{\mu\nu} ({\cal G} h)_{\mu\nu} &\equiv & -{1 \over 4} h_{\nu\rho;\mu} h^{\nu\rho;\mu} + {1 \over 2} h_{\mu\nu;\lambda} h^{\lambda\nu;\mu}  - {1 \over 2} h_{;\mu} h^{\mu\nu}_{\ \ ;\nu}+ {1 \over 4} h_{;\mu} h^{;\mu} \nonumber\\ &&  + {1 \over 2\ell^2}\left( h_{\mu\nu}h^{\mu\nu}-{1 \over 2}h^2 \right)
\end{eqnarray}
and we have used the shorthand notation $h\cdot h \equiv   h_{\mu\nu}h^{\mu\nu}-h^2$.
(Indices are raised and lowered with $\bar g_{\mu\nu}$.) The fluctuations $h_{\mu\nu}$ and $\rho_{\mu\nu}$ can be decoupled by a linear redefinition of fields,
\begin{equation}\label{dec}
\rho = h^{^{(0)}}-h^{^{(m)}}, \ \ \ \ \ \   h  = h^{^{(0)}} + \sigma h^{^{(m)}}
\end{equation}
The action becomes
\begin{equation}\label{Lin}
I ={1+\sigma \over 16\pi G} \int h^{^{(0)} \mu\nu} ( {\cal G}  h^{^{(0)}})_{\mu\nu} + {(1+\sigma)\sigma \over 16\pi G} \int\,\left[  h^{^{(m)} \mu\nu} {\cal G}( h^{^{(m)}})_{\mu\nu}  - {1 \over 4} m^2\,\Big( h^{^{(m)}\mu\nu} h^{^{(m)}}_{\ \ \mu\nu} - (h^{^{(m)}})^2 \Big) \right] .
\end{equation}
with
\begin{equation}\label{}
m^2 = {4\nu \over \ell^2} {1+\sigma \over \sigma }.
\end{equation}
In three dimensions $h^{^{(0)}\mu\nu}$ is trivial and can be omitted. On the other hand, the second term is exactly the Pauli-Fierz action for $h^{^{(m)}\mu\nu}$ describing a massive unitary spin 2 particle in three dimensions, with a new Newton constant ${1 \over G'}={(\sigma+1)\sigma \over G}$.   We have thus shown that at linearized level the action (\ref{I}) is fully equivalent to the action recently proposed in \cite{BHT}. We expect the two theories to differ at the non-linear level and at the quantum level.

The decoupling transformation (\ref{dec}) fails at $\sigma=-1$ where it becomes non-invertible. At this point, the action (\ref{I1}) can be expressed as
\begin{equation}\label{I2}
I[h_{\mu\nu},\rho_{\alpha\beta}] = {1 \over 16\pi G} \int \left( h^{\mu\nu}_{-} ({\cal G} h)_{+\mu\nu} - {\nu \over \ell^2} (h^{\mu\nu}_{+}h_{+\mu\nu} - h_{+}^2) \right)
\end{equation}
where $h_{\pm\mu\nu}=h_{\mu\nu}\pm\rho_{\mu\nu}$.  This action also arises in NMG theory \cite{BHT} for a particular value of the couplings. The field content is a propagating  massive vector field \cite{BHT}.

\section{Warped Black holes and critical point}

The action (\ref{I}) contains an anti-de Sitter background and thus it also contains black holes constructed as quotients of (\ref{bu}). This produces two identical $g_{\mu\nu}=f_{\mu\nu}$ three dimensional black holes \cite{BTZ,BHTZ} with the same mass and angular momentum. One can anticipate that these charges cannot be decoupled to have different values for each field. The reason is that $f_{\mu\nu}-g_{\mu\nu} = \rho_{\mu\nu}-h_{\mu\nu} = -(1+\sigma)h_{\mu\nu}^{^{(m)}}$ is short range (for generic masses\footnote{It would be very interesting to make a systematic study of `dangerous' masses with long-range fluctuations.}). Thus, asymptotically, both metrics approach each other.

The black holes constructed as quotients of (\ref{bu}) are asymptotically AdS, and thus have an asymptotic SO(2,2) symmetry. This symmetry is extended to the full conformal group \cite{BH} in the usual way.

Now, just as it happens in NMG, the action (\ref{I}) contains a richer spectrum of black holes. In this section we shall find black holes preserving SL(2,$\Re$)$\times \Re$, in a way that resembles the warped black holes discussed in \cite{Anninos:2008fx,Anninos:2009zi}. We also find a critical point in the space of couplings $\sigma,\nu$ where the full $SO(2,2)$ symmetry is restored. These black holes also obey generalized boundary conditions similar to the ones discussed in \cite{Compere:2009zj}.

Let us look for general solutions to the equations of motion with two commuting killing vectors $\partial /\partial t$ and $\partial /\partial \varphi$. Metrics with two commuting killing vectors are parameterized as
\begin{eqnarray}
  ds^2 &=& -f dt^2 + h dr^2  + r^2 d\varphi^2 + J dtd\varphi,  \nonumber\\
  df^2  &=& -X dt^2 + Y dr^2 + Z d\varphi^2 + L dt d\varphi +  U dt dr + V drd\varphi,
  \label{bh}
\end{eqnarray}
where we use the notation $df^2=f_{\mu\nu}dx^\mu dx^\nu$. All functions $f,h,J$ and $X,Y,Z,L,U,V$ depend only on $r$.  The functions $U,V$ can be eliminated from $df^2$ via a coordinate redefinition. However, the action is invariant only under diffeomorphisms acting simultaneously in $g_{\mu\nu}$ and $f_{\mu\nu}$. Thus, if $U,V$ are eliminated from $df^2$ they reappear in $ds^2$. In this sense they cannot be omitted from both $g_{\mu\nu}$ and $f_{\mu\nu}$ simultaneously.

The four-dimensional `Schwarzschild' problem (without angular momentum) for the action (\ref{I}) was solved in \cite{IshamStorey}. See \cite{bg1,bg2,bg3} for more recent discussions on this problem. In four dimensions there is only one extra function $U={1 \over 2}f_{rt}(r)$, due to the properties of the sphere (there are no invariant 1-forms). The solutions can be classified in two cases, $U=0$ or $U\neq 0$. As shown in \cite{IshamStorey}, the equations for the second case can be solved analytically, while the first case remains unsolved. In three dimensions there are two extra functions, $U,V$ and correspondingly there exists 4 cases ($U=0=V)$, $(U\neq 0 =V)$, $(U=0\neq V)$ and $(U\neq 0 \neq V)$. All four cases admits solutions. However, as in four dimensions, we were able to find an exact analytic solution only for the case $(U\neq 0 =V)$. We discuss its properties in this section.  In the following section we shall analyze the asymptotic structure of the equations and find asymptotic solutions for the other cases.

We plug the ansatz  (\ref{bh}) in the equations of  motion assuming $V=0$. The solution satisfying Brown-Henneaux conditions can be expressed as follows. First, the metric $g_{\mu\nu}$ is a  3d black hole \cite{BTZ,BHTZ} with a constant pre-factor,
\begin{eqnarray}\label{bhg}
   ds^2 = {1 \over 1 + 2\lambda\nu}\left[-\left( {r^2 \over \ell^2} - M_g \right) dt^2 + {dr^2 \over  {r^2 \over \ell^2}-M_g + {J_g^2 \over 4r^2}} + J_g dt d\varphi + r^2 d\varphi^2 \right].
\end{eqnarray}
(The radial coordinate appearing here differs from that in (\ref{bh}) by a constant rescaling.) The metric $f_{\mu\nu}$ is given by
\begin{eqnarray}\label{bhf}
df^{2} &=& {1 \over 2(1 + 2\lambda\nu)} \left[-\left({2r^2 \over \lambda^2\ell^2} - M_f \right) dt^2 + {  8r^2\ell^2( 2 r^4 \lambda^2 + 2r^2( M_f \lambda^2\ell^2 - \ell^2 M_g \lambda^2 - 2M_g \ell^2) + \ell^2 J_g^2)  dr^2 \over \lambda^2 ( 4r^4 - 4r^2 M_g \ell^2 + \ell^2 J_g^2)^2     }\right. \nonumber\\
 && \left. \  + J_g dt d\varphi -
 { 4\ell r \sqrt{ (r^2(2-\lambda^2) + \lambda^2\ell^2(M_g-M_f))(4r^2(2M_g-M_f\lambda^2) + J_g^2(\lambda^2-2)  )  }   dt dr \over  \lambda^2 (4r^4 - 4r^2M_g\ell^2 + J_g^2\ell^2)  } + r^2 d\varphi^2 \right]
\end{eqnarray}
(Here we use $8G=1$).  $M_g,J_g,M_f$ are arbitrary integration constants.
$M_g$ and $J_g$ are clearly the mass and angular momentum of the metric $g_{\mu\nu}$. $M_f$ plays the role of mass in the dual metric $f_{\mu\nu}$.  We are missing an independent charge $J_f$ for the metric $f_{\mu\nu}$.  This is due to our assumption $V=0$. We find the general asymptotic solution below. Note that there is no choice of $M_g,J_g,M_f$ which leads to (\ref{bu}).

The constant $\lambda$ is related to the couplings $\nu$ and $\sigma$ by the quadratic equation,
\begin{equation}\label{lambda}
2\nu\lambda^2+4\sigma\lambda\nu+\nu+\sigma=0
\end{equation}
This means that in principle there exists two solutions for each value of $\nu,\sigma$. We shall impose the extra condition $\lambda>0$, that eliminates one of them. This condition is necessary for the following reason. The equations of motion for the action (\ref{I}) contain $\sqrt{f/g}$. We assume that both volume elements are positive. Now, on the solution (\ref{bhf}) $\det {f_{\mu\nu}}$ has the simple expression,
\begin{equation}\label{}
\det( f_{\mu\nu} )= -{r^2 \over 4\lambda^2( 1+2\lambda\nu )^3}.
\end{equation}
The factor $1+2\lambda\nu$ must be positive (see, for example, (\ref{bhg})). Then $\sqrt{-f}>0$ requires $\lambda>0$. If $\lambda>0$, then from (\ref{lambda}) we see that $\sigma$ or $\nu$ must be negative. This is not a problem because, on the one hand, gravitons may have negative masses on AdS, and on the other, $\sigma$ could be negative without spoiling unitarity (wrong-right sign) \cite{BHT}, as it is clear from (\ref{Lin}).

Note that $f_{tr} \sim {\cal O}(1/r)$ at space-like infinity. This is just enough to allow a conformal structure. We discuss this in detail below (see \cite{Compere:2009zj} for a related  analysis). There are solutions to the action (\ref{I}) in which $U \sim {\cal O}(1)$, but we discard them because they do not have a good AdS structure.

Let us investigate the asymptotic behavior of this solution. We keep here only the dominant terms which already have the information we need. For very large $r$ the solution approaches
\begin{eqnarray}\label{}
ds^2 & \simeq & {1 \over 1 + 2\lambda\nu}\left[ {\ell^2 dr^2 \over r^2}  + r^2 d\varphi^2 - {r^2 \over \ell^2} dt^2 \right] \label{abhg}\\
df^2 &\simeq & {1 \over 2(1 + 2\lambda\nu)}\left[ {\ell^2 dr^2 \over r^2}  + r^2 d\varphi^2 - {2 \over \lambda^2} {r^2 \over \ell^2} dt^2   \right] \label{abhf}
\end{eqnarray}
We conclude that for generic values of $\lambda$ this configuration is not asymptotically AdS. This point requires some explanation. Both metrics (\ref{abhg}) and (\ref{abhf}) are asymptotically AdS, but with different speeds of light because the coefficients of $dt^2$ are different. Due to the factor ${2 \over \lambda^2}$, the metric $df^2$ appears warped with respect to $ds^2$. Hence, even though each metric has 6 Killing vectors, only four of them are common to both metrics.

The generators which leave both (\ref{abhg}) and (\ref{abhf}) invariant are constant time translations $t\rightarrow t+a_0$, plus the SL(2,$\Re)$ isometries of the Euclidean 2-dimensional AdS$_2$ factor ${\ell^2 dr^2 \over r^2}  + r^2 d\varphi^2$ which is common to both metrics. The full residual group is then SL(2,$\Re) \times \Re$.  This resembles very much the warped solutions of \cite{Anninos:2008fx} where the symmetry is broken by a constant factor multiplying the fiber when one writes the AdS$_3$ metric as a U(1) fibration over AdS$_{2}$.

A critical case occurs for the particular value $\lambda^2=2$, where both asymptotic metrics do become equal. For this particular value of $\lambda$ the solution is asymptotically AdS$_3$ and a direct conformal structure can be read off (see below).
Now, recall that $\lambda$ is not an arbitrary constant but given by (\ref{lambda}).
The particular value $\lambda^2=2$ can occur if and only if the parameters $\nu,\sigma$  are related by
\begin{equation}\label{crit}
5\nu + \sigma + 4 \sqrt{2} \sigma \nu =0.
\end{equation}
This equation defines a critical line in the space of couplings where the asymptotic symmetry SL(2,$\Re)\times \Re$ is enhanced to SO(2,2) and, in fact, the full conformal group.  Condition (\ref{crit}) is not an artifact of the particular class of black holes with $V=0$. We find below the full asymptotic solution to the equations of motion and recover the same condition.

At $\lambda^2=2$ the space of solutions (\ref{bhg}) and (\ref{bhf}) also contains an SO(2,2) invariant ground state. For $M_g=M_f=-1$ and $J_g=0$, both metrics (\ref{bhg}) and (\ref{bhf}) become proportional to global AdS space,
\begin{equation}\label{bv}
d s^2 =  {1 \over 1 + 2\sqrt{2}\nu}\left[ - \left( 1 +  {r^2 \over \ell^2} \right) dt^2 + { dr^2 \over 1 + {r^2 \over \ell^2} }+ r^2 d\varphi^2 \right], \ \ \ \ \ \mbox{and} \ \ \ \ \  df^{2}={1 \over 2} d s^2.
\end{equation}
This state then has six Killing vectors generating the group $SO(2,2)$. If (\ref{crit}) did not hold, the space of solutions would not have an SO(2,2)-invariant state.

The black hole (\ref{bhg}) is clearly a quotient of the AdS ground state (\ref{bv}). This is less clear for the metric (\ref{bhf}), although one can check that it also has constant curvature. In fact, (\ref{bhf}) can be put in the form (\ref{bhg}) by a change of coordinates. A similar property also holds in four dimensions \cite{IshamStorey}: Both the metric $g_{\mu\nu}$ and $f_{\mu\nu}$ are isometric to the Schwarzschild solution, but they are in different coordinate systems. As a matter of fact, the generic 4d solutions found in \cite{IshamStorey} are not asymptotically AdS. If one imposes AdS fall off conditions on \cite{IshamStorey} then a condition similar to (\ref{crit}) arises in four dimensions as well.

The global constant factors appearing in (\ref{bhg}) and (\ref{bhf}) ensure that these solutions span a different sector of the theory, not related to the background (\ref{bu}) and its quotients. In fact, the solutions built from the background (\ref{bu}) are transparent to the potential, while (\ref{bhg}) and (\ref{bhf}) depend explicitly on $\nu$.

The outcome of this discussion is that the action (\ref{I}), with the couplings related by (\ref{crit}), has two different AdS groundstates. Each of them have `excited states', or black holes. Black holes on (\ref{bu}) are characterized by only two charges $M,J$. Black holes on (\ref{bv}) have three charges. Actually the black holes on (\ref{bv}) have more charges which are not seen in the above solution because we have assumed $V=0$ in the ansatz (\ref{bh}). In the following section we study the asymptotic structure of black holes on (\ref{bv}) in general, and show that the solutions are characterized by more parameters.

Symmetry breaking is at the heart of bigravity \cite{Berezhiani:2007zf,BlasDeffayetGarriga} and one may wonder whether demanding AdS asymptotics restricts the power of the theory. We hope to come back to this issue elsewhere.

\section{Asymptotics and conformal Structures}

We have proved that the action (\ref{I}), with the couplings related by (\ref{crit}),  has two AdS$_3$ phases. A first AdS background is given by (\ref{bu}).  A second class of AdS backgrounds is given in (\ref{bv}). Without going through the details of the previous section we can re-derive the main results on the backgrounds as follows.

We shall use a notation appropriate to the conformal structure. First, we introduce two new coordinates  $z,\bar z$ related to the Schwarzschild coordinates $t,\varphi$ as
\begin{equation}\label{}
z = {t \over \ell} + \varphi, \ \ \ \ \ \  \bar z = -{t \over \ell} + \varphi.
\end{equation}
The asymptotic form of the metric (\ref{bu}) in these coordinates reads
\begin{equation}\label{adsb}
ds^2 \sim  {\ell^2 dr^2 \over r^2} + r^2 dz d\bar z .
\end{equation}
Consider the following family of AdS backgrounds,
\begin{equation}\label{other}
g_{\mu\nu} = \beta\, \bar g_{\mu\nu}, \ \ \ \ \ \  f_{\mu\nu} = \gamma\, \bar g_{\mu\nu},
\end{equation}
where $\beta$ and $\gamma$ are constants and $\bar g_{\mu\nu}$ is given in (\ref{bu}). The metric $\bar g_{\mu\nu}$ is SO(2,2) invariant. Since $\beta$ and $\gamma$ are constants, the configuration (\ref{other}) is also SO(2,2) invariant.

We plug (\ref{other}) into the equations of motion and find the following two conditions for $\beta$ and $\gamma$,
\begin{eqnarray}\label{beta}
4\nu\beta(\beta-\gamma) + \sqrt{\beta\gamma}(\beta-1)&=&0, \nonumber\\
 \gamma^2(\sigma+3\nu) -\gamma(\sigma+2\beta \nu) - \nu\beta^2&=&0.
\end{eqnarray}
The background (\ref{bu}) corresponds to $\beta=\gamma=1$, which is a solution to this system, but there are other solutions.

Among all the solutions contained in (\ref{beta}), we are particularly interested in the critical theory satisfying (\ref{crit}) because it contains SO(2,2) black holes. If (\ref{crit}) holds then (\ref{beta})
has the solution
\begin{equation}\label{gamma}
\beta = {1 \over 1 + 2\sqrt{2}\nu }, \ \ \ \ \ \ \gamma = {1 \over 2(1 + 2\sqrt{2}\nu) }
\end{equation}
and we recognize the background (\ref{bv}) and its associated black holes (\ref{bhg},\ref{bhf}).

Our goal now is to study linearized asymptotic fluctuations on the backgrounds (\ref{other}), where (\ref{crit}) is satisfied and $\gamma$ and $\beta$ are given by (\ref{beta}). Our main result is that the following fields
\begin{eqnarray}
ds^2 &\sim& \beta \ell^2\left( {dr^2\over r^2} + {r^2 \over \ell^2} dz d\bar z + T(z) dz^2 + \bar T(\bar z)d\bar z^2 \right) + \cdots  \label{newa} \\
df^2 &\sim& \gamma\ell^2 \left( {\ell^2dr^2 \over r^2} + {r^2 \over \ell^2} dz d\bar z  + Q(z)dz^2 + \bar Q(\bar z) d\bar z^2 + {2dr \over r} ( P(z) dz + \bar P(\bar z) d\bar z )\right)   + \cdots   \nonumber
\end{eqnarray}
satisfy the asymptotic equations, where $T,Q,P$ and $\bar T,\bar Q,\bar P$ are arbitrary functions of their arguments.  The functions $P$ and $\bar P$ appearing in (\ref{newa}) can be set to zero by a simple (trivial, zero charge) redefinition of coordinates. For example, $P$ can be eliminated by $\bar z = \bar z' + {\ell^2 \over 2}P(z)/r^2$, with the effect of redefining $Q$.  However, if we eliminate $P$ from $f_{\mu\nu}$, it reappears in the metric $g_{\mu\nu}$. In this sense, $P$ and $\bar P$ are physical. Note also that the black hole (\ref{bhf}) contains a contribution of this form.  The fluctuations $T$ and $\bar T$ are just the usual Brown-Henneaux fields.

This asymptotic solution has four charges, as promised. A crucial point is that (\ref{newa}) is a solution if and only if (\ref{crit}) is satisfied. This has several consequences. The action (\ref{I}) describes massive gravitons and one may ask why terms of the form $r^{\alpha}$, where $\alpha$ is some non-integer function of the mass, have not appeared in the asymptotic solution (\ref{newa}).  The reason is simple. The action (\ref{I}) describes massive gravitons when expanded around the background (\ref{bu}). The solutions (\ref{newa}) are fluctuations on a different background (\ref{other}), whose spectrum is different. It is easy to prove that linearization on the background (\ref{other}) does not a give Pauli-Fierz theory but instead a `mass' term of the form $(h^{\mu}_{\ \mu})^2$. This does not really provide a mass for the graviton, and explains why the expansion (\ref{newa}) has no $r^\alpha$ terms.

The solutions (\ref{newa}) are the most general fields consistent with Brown-Henneaux transformations.  That is, under the coordinate redefinitions (Brown-Henneaux diffeomorphisms),
\begin{eqnarray}\label{con}
z' &=& z+ \epsilon(z) - {1 \over 2 }{\ell^2 \over r^2} \bar \partial^2 \bar\epsilon(\bar z), \\
\bar z' &=&  \bar z + \bar \epsilon(\bar z) - {1 \over 2 }{\ell^2 \over r^2} \partial^2 \epsilon(z), \\
r' &=& r-{r \over 2} (\partial \epsilon(z)+  \bar\partial\bar \epsilon(\bar z)  ),
\end{eqnarray}
where $\epsilon(z)$ and $\bar\epsilon(\bar z)$ are arbitrary functions of their arguments, these metrics transform among themselves with new fields $T',Q',P'$ and $\bar T',\bar Q',\bar P'$. Let $\delta T \equiv T'(z)-T(z)$ and the same for the other fields. Plugging (\ref{con}) into (\ref{newa}) one obtains the transformations,
\begin{eqnarray}\label{dQ}
\delta T &=& -\epsilon \partial T - 2\partial \epsilon T + {1 \over 2} \partial^3 \epsilon, \nonumber\\
  \delta Q &=& -\epsilon \partial Q - 2\partial \epsilon Q + P \partial^2 \epsilon + {1 \over 2} \partial^3 \epsilon, \\
\delta P &=& -\epsilon \partial P - \partial\epsilon P,
\end{eqnarray}
and corresponding equations for the (psudo) anti-holomorphic fields.  $T$ is the usual Brown-Henneaux Virasoro field transforming with weight $(2,0)$ and a central term.
The new field $P$ has conformal weight $1$, as expected since it appears in the metric as a 1-form $P dz$. However,  $Q$ does not transform with definite conformal weight, but has a contribution from $Q$. Let us point out that the combination $ \hat Q \equiv Q + \partial P$ does transform correctly with $h=2$ and a central term. We now prove that this combination is in fact what shows up in the charge that generates conformal transformations.

Since the action is the sum of two Einstein-Hilbert actions (the interaction does not have derivatives), the conserved charge associated to the asymptotic symmetries is simply the sum of two ADM functionals. The total charge $J$ (see, for example, \cite{BH}) is,
\begin{equation}\label{}
J = J_{ADM}[g] + \sigma \, J_{ADM}[f].
\end{equation}
By direct calculation we find the total charge to be
\begin{equation}\label{J}
J(\epsilon) = {1 \over 4G}\int {d\phi \over 2\pi} \epsilon \left(\sqrt{\beta} T + \sigma\sqrt{\gamma}( Q+\partial P)\right).
\end{equation}
As we have suspected before hand, the relevant Virasoro charge in the $f-$sector is not $Q$ but $Q + \partial P$.  The total charge is thus the sum of two Virasoro operators transforming correctly under the conformal group.

The central charge can be computed directly from the transformations (\ref{dQ}), the knowledge of the charge (\ref{J}), and the fact that $\delta_{\rho}J(\epsilon)=\{J(\rho),J(\epsilon)\}$\cite{Brown:1986ed}. We can arrive at the desired result in a quicker way as follows. We know that for one metric on the background (\ref{bu}) the central charge is $3\ell/2G$ \cite{BH}.

The background (\ref{other}) differ from (\ref{bu}) by the factors $\beta$ and $\gamma$ in each metric. Note that if $g_{\mu\nu}=\beta \bar g_{\mu\nu}$ then $\sqrt{g}g^{\mu\nu}R_{\mu\nu}(g)=\sqrt{\beta} \sqrt{\bar g}\bar g^{\mu\nu} R_{\mu\nu}(\bar g)$ (in three dimensions). Putting everything together the total central charge is
\begin{eqnarray}\label{}
c &=& {3\ell \over 2G} ( \sqrt{\beta} + \sigma\sqrt{\gamma} ) \nonumber\\
 &=&  {3\ell \over 2G}{1 \over \sqrt{1 + 2\sqrt{2}\nu  }} \Big(1  + {\sigma \over \sqrt{2}} \Big)
\end{eqnarray}
where in the second line we have used (\ref{gamma}). Recall that $\sigma$ can be written in terms of $\nu$ using using (\ref{crit}). These conformal models are parameterized by a single real constant $\nu$.

This central charge can be compared with the CFT associated with the original background (\ref{bu}). In this case $g_{\mu\nu}$ contributes to $c$ with $3\ell/2G$ and $f_{\mu\nu}$ contributes with $3\ell/2G \times \sigma$ (see the action (\ref{I})). The total central  charge is
\begin{equation}\label{}
c_0 = {3\ell \over 2G}(1+\sigma)
\end{equation}

The backgrounds (\ref{other}) make sense provided $\gamma$ and $\beta$ are positive. This implies that $\nu$ must lie in the range $-{1\over 2\sqrt{2}}<\nu<\infty.$  In this range both $c$ and $c_0$ are positive and it can be checked that $c_0 > c$, for all the allowed range of $\nu$. A solution interpolating both AdS vacua flowing from $c_0$ (UV) to $c$ (IR) may exist.

\section{Acknowledgements}

M.B. would like to thank G. Barnich for useful discussions and warm hospitality at ULB where part of this work for done. M.B. was partially supported by Alma grant \#  31080001; Fondecyt Grant \#1060648; and the J.S. Guggenheim Memorial Foundation. The work of S.T. was supported by the German-Israeli Project cooperation (DIP) and the German-Israeli Foundation (GIF). The authors would like to thank ESI, Vienna, for hospitality, and the organizers of the ``ESI Workshop on 3D Gravity'' where this work was initiated, for the great atmosphere provided.


 \end{document}